WILEY-VCH

**Article type: Full Paper**

**Single-mode lasing from colloidal water-soluble CdSe/CdS quantum dot-in-rods**

*Francesco Di Stasio, Joel Q. Grim, Vladimir Lesnyak, Prachi Rastogi, Liberato Manna, Iwan Moreels* and Roman Krahne**

Dr. F. Di Stasio, Dr. J.Q. Grim, Dr. V. Lesnyak, P. Rastogi, Prof. L. Manna, Dr. I. Moreels, Prof. R. Krahne
Istituto Italiano di Tecnologia, Via Morego 30, IT-16163 Genoa, Italy
E-mail: Roman.Krahne@iit.it, Iwan.Moreels@iit.it

*This is the pre-peer reviewed version of the following article Francesco Di Stasio, Joel Q. Grim, Vladimir Lesnyak, Prachi Rastogi, Liberato Manna, Iwan Moreels, and Roman Krahne, Single-mode lasing from colloidal water-soluble CdSe/CdS quantum dot-in-rods, Small 11, 1328–1334, March 18, 2015, which has been published in final form at https://onlinelibrary.wiley.com/doi/full/10.1002/smll.201402527. This article may be used for non-commercial purposes in accordance with Wiley Terms and Conditions for Self-Archiving.*

Abstract:

Core-shell CdSe/CdS nanocrystals are a very promising material for light emitting applications. Their solution-phase synthesis is based on surface-stabilizing ligands that makes them soluble in organic solvents, like toluene or chloroform. However, solubility of these materials in water provides many advantages, such as additional process routes and easier handling. So far, solubilization of CdSe/CdS nanocrystals in water that avoids detrimental effects on the luminescent properties, poses a major challenge. This work demonstrates how core-shell CdSe/CdS quantum dot-in-rods can be transferred into water using a ligand exchange method employing mercaptopropionic acid (MPA). Key to maintaining the light-emitting properties is an enlarged CdS rod diameter, which prevents potential surface defects formed during the ligand exchange from affecting the photophysics of the dot-in-rods. Films made from water-soluble dot-in-rods show amplified spontaneous emission (ASE) with a similar threshold (130 μJ/cm$^2$) as the pristine material (115 μJ/cm$^2$). To demonstrate feasibility for lasing applications, self-assembled micro-lasers are fabricated via the "coffee-ring effect" that display single-mode operation and a very low threshold of ~10 μJ/cm$^2$. The performance of these micro-lasers is enhanced by the small size of MPA ligands, enabling a high packing density of the dot-in-rods.

## 1. Introduction

Colloidal semiconductor nanocrystals (NCs)[1] have attracted increasing attention in the last two decades due to their potential for solution-processed and flexible optoelectronics. A variety of colloidal NCs based light-emitting diodes,[2-6] lasers[7-14] and non-linear optical absorbers[15] have been demonstrated. The main attractiveness of colloidal NCs for solution-processed optoelectronics is their versatile chemical synthesis, which allows their optical properties to be tailored by controlling size, shape and composition. Furthermore, colloidal NCs based on core/shell heterostructures allow for additional tuning of their photophysics, by enabling control over quantum confinement via formation of type-I, type-II or "quasi-type II" hetero-junctions.[16]

In recent years, colloidal CdSe/CdS NCs have been widely investigated, and many different architectures have been synthesized and studied. In particular, in CdSe/CdS structures with a quasi-type II hetero-junction, the holes are localized within the CdSe core, while the electrons are more delocalized due to the small conduction band offset.[17] The resulting larger exciton volume leads to a significant decrease of the Auger recombination rate. CdSe/CdS asymmetric structures such as quantum dot-in-rods (QDRs) synthesized by seeded-growth[18] further benefit from reduced optical gain threshold,[7, 19] owing to enhanced optical absorption by the CdS rod,[20] as well as the ability to form densely packed ordered multilayer films.[5, 21] Moreover, nearly temperature-independent amplified spontaneous emission (ASE)[20] has been recently demonstrated for this class of colloidal

NCs. All these properties have stimulated the study and development of CdSe/CdS QDRs for laser applications.[9, 10, 20, 22]

From a device point of view, water soluble NCs are desirable since they enable the fabrication of architectures based on materials processable in orthogonal solvents (e.g. multi-layer structures produced via sequential spin-coating),[23] as well as the exploitation of nanostructures fabricated with organic-soluble conjugated[24] or saturated polymers.[25] However, obtaining water-soluble CdSe/CdS NCs possessing similar light-emitting properties as organic soluble ones has posed a major challenge. This is due to the creation of additional photoluminescence (PL) quenching channels arising from surface defects after ligand exchange, which leads to a decrease of photoluminescence quantum yield (PLQY).[26] To overcome this issue, different techniques based on encapsulation of NCs into water-soluble shells have been developed as, for example, poly(maleic anhydride-alt-1-octadecene)[27] or phospholipid block–copolymer micelles.[28] Yet, all these methods require multiple steps and yield NCs encapsulated in a thick organic shell, which hinders the formation of densely packed films (e.g. Au nanoparticles functionalized with poly(maleic anhydride-alt-1-octadecene) show an increase in radius from 7.6 to 13.6 nm after the encapsulation)[27] and may require a control of the pH to prevent NCs precipitation.

In 2003, Wuister et al.[29] successfully transferred CdTe NCs into water employing a ligand exchange reaction based on mercaptopropionic acid (MPA, forming negatively charged NCs), and they demonstrated PLQY as high as 60% at room temperature. However, the MPA based ligand exchange method applied to CdSe/CdS heterostructures yielded low PLQY.[29] Here, we report the optical properties of highly luminescent CdSe/CdS QDRs transferred into water using the MPA ligand exchange. We show that by increasing the CdS rod diameter (with respect to the CdSe core size), it is possible to maintain the light-emitting properties of the pristine heterostructures. This is demonstrated by similar PLQY (50 ± 5% for pristine and MPA capped QDRs) and PL dynamics. The larger CdS rod prevents surface defects formed during the reaction to affect the photophysics of the material. Additionally, MPA as a surface ligand leads to shorter inter-QDRs distances in assemblies than the pristine octadecylphosphonic acid (ODPA), allowing the fabrication of higher density films. The observation of a similar ASE threshold (from the emissive states of the CdSe core) from films made of MPA capped QDRs as for the pristine ODPA QDRs, and laser emission from self-assembled structures further motivates the use of water-soluble QDRs as gain media in laser architectures. The material design criteria proposed here (i.e. employing a CdS rod with a substantially larger diameter than the CdSe core) can be extended to other classes of CdSe/CdS heterostructures such as core-shell nanocrystals[16, 30, 31] (including the so-called "giant-shell NCs"[3, 4, 32, 33]), rod-in-rods[34] and colloidal quantum wells.[35]

## 2. Results and discussion

**Figure 1a** shows a scheme of the core-shell QDRs used in the study. The average dimensions of the CdSe/CdS QDRs were extracted from transmission electron microscope (TEM) image analysis. The QDRs were synthesized via seeded-growth[18] with a rod diameter of 6.6 nm, which is 35% larger than the CdSe core (4.3 nm), and results in a CdS shell thickness $a_{CdS} = 1.15$nm (see Figure 1a). The seeded growth proceeds typically in within minutes, and leads to well defined core-shell structures as was demonstrated by mean dilatation mapping of high resolution transmission electron microscopy (TEM) images in ref.[18]. Anion diffusion between the CdSe core and the CdS shell, and associated interface alloying can therefore assumed to be limited.

As-synthesized QDRs were capped with ODPA (structure shown in Figure 1b), which grants solubility in non-polar solvents. The water transfer (see scheme in Figure 1b) was carried out by adding 100 µL of toluene solution of ODPA capped QDRs to 1-2 mL of 0.1 M MPA and 0.12 M KOH dissolved in methanol, depending on the initial concentration of QDRs in the toluene solution (varying from 5 to 10 µM). The mixture was left stirring for 20 minutes to complete the ligand exchange, and centrifuged and purified through precipitation using isopropanol. The QDRs with MPA capping were then dispersed in deionized water (Figure1b).

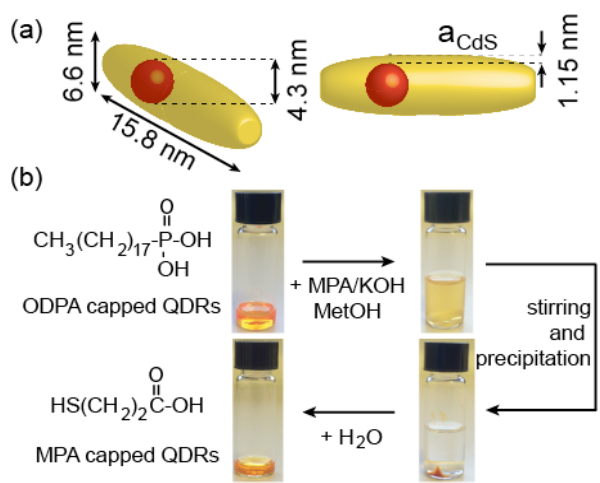

**Figure 1.** (a) Illustration of a CdSe/CdS QDR used in this study with its average dimensions determined *via* TEM analysis. The shell thickness $a_{CdS}$ at the core is depicted in the cross section on the right. The CdS rod diameter is 35% larger than the CdSe core corresponding to a CdS shell thickness $a_{CdS}$ of 1.15 nm. (b) Scheme of the ligand exchange reaction employed for the water-solubilization of the QDRs.

In **Figure 2a** and 2b TEM micrographs of ODPA (a) and MPA (b) capped QDRs are presented (additional micrographs are shown in Figure SI1). Notably, a different packing is observed for MPA and ODPA capped QDRs: we can tentatively assign this effect to differences in the solvent/substrate interactions, solvent surface tension (ODPA capped QDRs were dispersed in toluene while water was used as a solvent for MPA capped QDRs), as well as the polar character of MPA which will affect the interaction between QDRs. As expected from the chain length of the two ligands (Figure 1b), we obtained average inter-QDRs distances from TEM images of 2.5 ± 0.6 nm and 1.6 ± 0.5 nm for ODPA and MPA capped QDRs respectively.[36] The smaller size of MPA ligands leads to a larger CdSe/CdS volume fraction ($V_f = V_{CdSe\text{-}CdS} / V_{QDR}$) of ~ 0.41 compared to ~ 0.27 for ODPA. As a result, the smaller MPA ligands should enable the formation of higher density films.

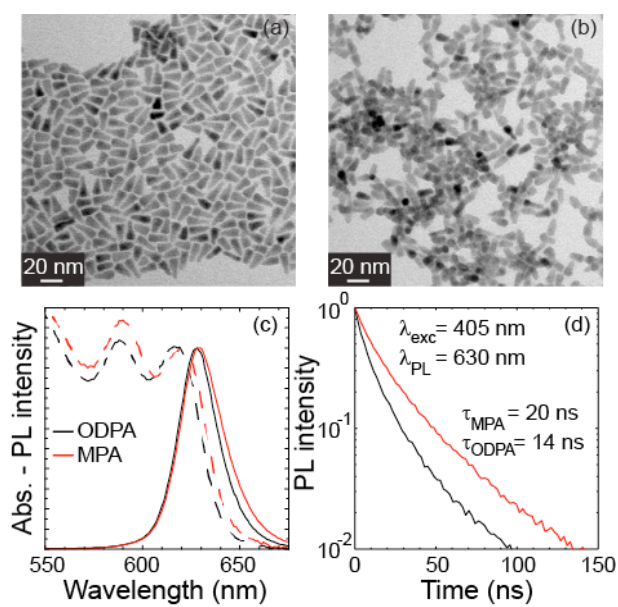

**Figure 2.** TEM images of (a) ODPA and (b) MPA capped QDRs. The different QDR packing and arrangement observed may be explained by different evaporation conditions or liquid-substrate interactions of the two solvents used (toluene for ODPA capped QDRs and water for MPA capped ones) as well as the polar character of the MPA ligands. (c) Normalized PL and optical absorption spectra for ODPA (black) and MPA (red) capped QDRs (color coding is the same for all panels). A slight red shift and broadening of 2 nm of the PL peak is observed for MPA capped QDRs (PL FWHM = 23 nm and 26 nm for ODPA and MPA capped QDRs, respectively). (d) PL decays of ODPA and MPA capped QDRs measured at $\lambda$ = 630 nm, excited with a pulsed laser diode ($\lambda$ = 405 nm, pulse width = 50 ps). An increase in the effective PL lifetime ($\tau$) of MPA capped with respect to OPDA capped QDRs is observed (from $\tau$ = 14 ns for ODPA capped QDRS to $\tau$ = 20 ns for MPA capped ones). The effective PL lifetimes reported in (d) are the weighted average of the time-constants and intensities used in the fitting procedure of the PL decay.

Generally, in QDRs the band-edge exciton transition energy is given by the size of the CdSe core, with an additional red-shift induced by the increased diameter in the final CdSe/CdS QDR due to electron delocalization.[11,17, 37] In Figure 2c we report the optical absorption spectra of QDRs before (ODPA capped - black line) and after (MPA capped - red line) transfer into aqueous solution recorded from diluted solutions, where we observe a further small, ligand-induced, red-shift of ~ 2 nm after the exchange of ODPA with MPA, i.e. absorption peaks for ODPA at 588 and 617 nm shift to 590 and 619 nm for MPA, respectively. A similar red-shift is also found for the PL peak, from 627 nm to 629 nm. In both the optical absorption and PL spectra, the red-shift is accompanied by a slight broadening, which in the PL spectrum leads to a full-width-half-maximum (FWHM) increase from $\sim$ 23 nm to $\sim$ 26 nm. The PL red-shift and broadening can be explained by a minor increase of the CdS rod diameter after the ligand exchange reaction, caused by an additional sulfur layer from the MPA (see chemical structure in Figure1b), which slightly decreases the wave function confinement. Hence, different ligand densities on the QDR surfaces may introduce a slight heterogeneous broadening.

The temporal evolution of the PL measured in a spectral range of 10 nm centered around the PL maximum at $\sim$ 630 nm (see Figure 2d) shows an increase of the PL lifetime ($\tau$) from $\sim$ 14 ns to $\sim$ 20 ns ($\sim$ 42% increase) following the ligand exchange reaction. The increase of PL lifetime mostly results from changes in the dielectric screening caused by the different solvents used: chloroform for the ODPA, and water for the MPA capped QDRs. In fact, by applying the Maxwell-Garnett effective medium theory[38, 39] (see SI), we can assign the increase in PL lifetime predominantly to variations of the local field factor[40] that affect the radiative rate via Fermi's golden rule. Considering that the addition of 1 monolayer of CdS already leads to a red-shift of 7.3 nm in spherical CdSe/CdS NCs,[41] we can assume that the sulfur atoms from the ligand shell, while inducing a small 2 nm red-shift of the PL (as observed in Figure 2c), will not lead to a substantial decrease of the e-h overlap. PLQY measurements carried out in an integrating sphere show similar values of 50 ± 5 % for both ODPA and MPA capped QDRs. The same ligand exchange reaction applied to QDRs possessing a CdS rod with diameter of 5.4 nm and a CdSe core of 5.3 nm diameter (i.e., $a_{CdS}$ = 0.05nm see Figure SI2) yields a sharp decrease of the PLQY from 50 ± 5% to 10 ± 1%, which can be rationalized by a strong impact of surface defects formed during the ligand exchange on the optical properties of the material. These observations indicate that the crucial parameter controlling the impact of potential surface defects is the distance between the CdSe core and the CdS rod surface, and that in our case a $a_{CdS}$ of about 1.15 nm is sufficient to preserve the PL properties.

Organic soluble CdSe/CdS QDRs have been extensively investigated as gain material, and the ASE threshold typically lies in the range of 150 μJ/cm$^2$.[11, 20] For water soluble QDRs to be a practical gain material for lasers, their ASE threshold should be similar to that of organic soluble QDRs. To this aim, we investigated the emission properties of QDR films under fs-pulsed excitation. **Figures 3**a and 3b show the streak camera images of ODPA (a) and MPA (b) capped QDRs films on soda-lime glass, recorded at fluences below the ASE threshold (50 μJ/cm$^2$). By increasing the optical pump fluence, an ASE peak appears at 628 nm and 629 nm for ODPA and MPA capped QDRs, respectively. At higher pump fluence (425 μJ/cm$^2$) the ASE peak is slightly red-shifted compared to the PL: 630 and 632 nm for ODPA and MPA capped QDRs, respectively (Figures 3c and 3d). The 3 nm red-shift of the ASE peak can be explained by attractive exciton-exciton interactions in the CdSe core and its surrounding, suggesting a type-I like behavior of the QDRs investigated.[22]

More importantly, the streak camera images show a spectral narrowing (in both cases a FWHM of about 8 nm is observed) and a drastically shortened temporal decay, clear fingerprints of ASE.[20] By integrating the streak camera images at increasing pump fluence over the full time range of 8 ns, we observe that the ASE appears for pump fluences above 115 μJ/cm$^2$ for ODPA capped QDRs and 130 μJ/cm$^2$ for MPA capped QDRs (Figures 3e and 3f). The evolution of PL and ASE for ODPA and MPA capped QDRs with increasing pump fluence is shown in Figures 3g and 3h. PL and ASE intensities were extracted from the time-integrated spectra using a sum of Gaussian peaks. The ASE peak is initially observed at the PL saturation onset, yielding a threshold of ~ 100 μJ/cm$^2$ (Figures 3g and 3h), well within the range of previously reported values for QDRs.[11, 20] This ASE threshold is also close to other comparable NCs based gain materials as, for example, CdSe/ZnCdS colloidal quantum dots.[12]

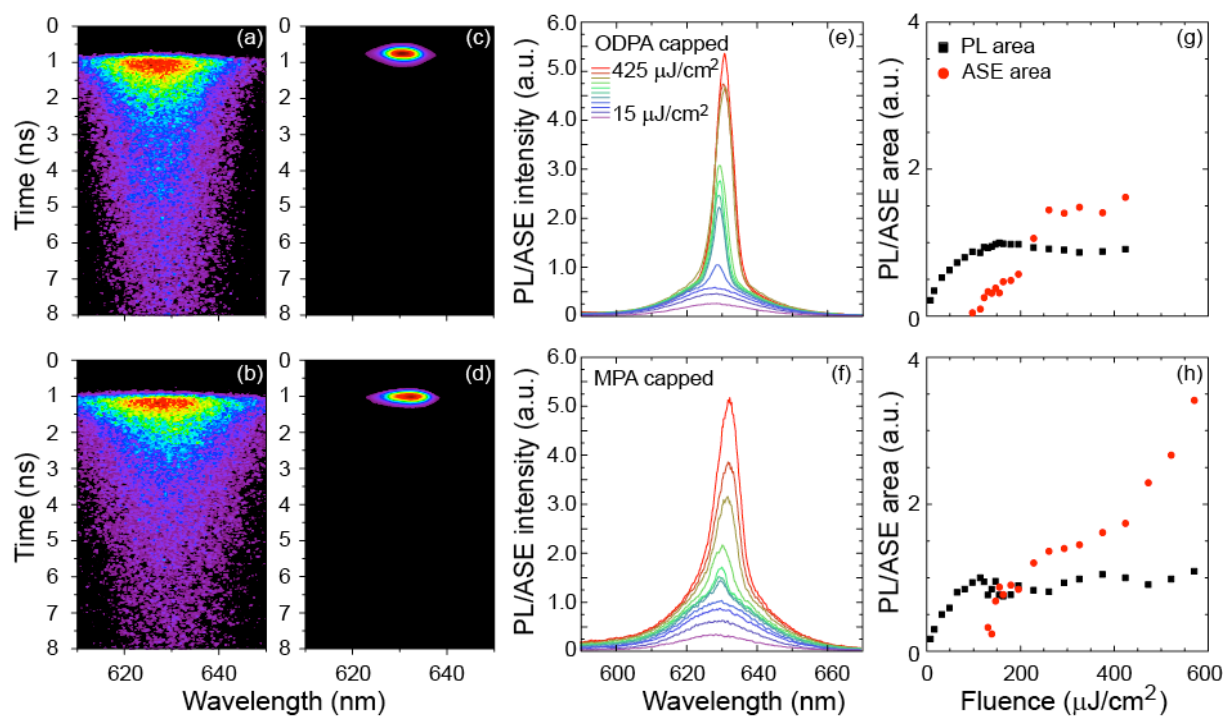

**Figure 3.** Emission properties of ODPA (top panels) and MPA (bottom panels) capped QDRs under fs-pulsed excitation at different pump fluences. Streak camera images recorded below (∼ 50 μJ/cm², a and b) and above (∼ 425 μJ/cm², c and d) ASE threshold for ODPA (a, c) and MPA (b, d) capped QDRs films, respectively. As expected, emission lifetime is shortened to the camera time-resolution when excitation fluence is above the ASE threshold. (e, f) Emission spectra obtained by integrating the streak camera images (over 8 ns) measured at different excitation fluences (from 15 μJ/cm² up to 425 μJ/cm²) for ODPA (e) and MPA (f) capped QDRs films. The ASE peak is initially observed at 628 nm, 115 μJ/cm² for ODPA capped QDRs and at 629 nm, 130 μJ/cm² for MPA capped QDRs, respectively. (g,h) PL and ASE intensity at increasing pump fluence for ODPA (g) and MPA (h) capped QDRs films. The intensities have been obtained from a multi-peak fit using a sum of two Gaussians to fit the ASE and PL. Extrapolation of ASE intensities at increasing pump fluences lead to an ASE threshold of ∼ 100 μJ/cm² for both ODPA and MPA capped QDRs, in good agreement with the values estimated from the time-integrated emission spectra (e, f).

To demonstrate the potential of MPA capped CdSe/CdS QDRs as gain-medium for lasers, we have fabricated self-assembled micro-lasers based on the "coffee-ring effect".[8, 42] By depositing a drop of typically 10 nL on planar glass surfaces, rings of colloidal particles are formed due to the pinning of the external contact line of the droplet on the substrate. While the droplet evaporates, an outward liquid flow maintains the droplet diameter, carrying the colloidal particles to the droplet border, forming a circular deposit.[42, 43] When the deposit height exceeds the height of the liquid surface, the droplet detaches, thus leaving a solid ring-like structure.[44] Typical coffee-ring micro-lasers obtained from our water soluble QDRs had a diameter around 500 μm (**Figure 4a**). Atomic force microscopy (AFM) images of the top-right section of the micro-laser (Figure 4b) show that the ring formed by the CdSe/CdS QDRs deposit has a total width of about 10 μm (at the base of the deposit), a FWHM of 6.8 μm and a height of 370 nm. The height is mainly determined by the concentration of the QDRs solution (MPA capped QDRs water solutions for coffee-ring fabrication were diluted to ~ 0.1 μM), whereas the droplet size (i.e. volume of solution deposited on the substrate), cleanliness of the substrate and contact angle can affect its shape. Coffee-rings obtained from QDRs in toluene solutions showed a preferred alignment in the packing along the tangential direction of the coffee-ring.[8] However, in the coffee-rings from the MPA capped QDRs water solutions discussed here we could not identify any specific packing order (see Figure S4 in the SI).

The fabrication of these structures using water-based QDRs can be performed in ambient conditions (i.e. no chemical hood is required as it is the case for NCs in organic solvents like chloroform), and presents several other advantages compared to organic solvents.[8, 9, 43] The high boiling point and slow evaporation of water should allow a high percentage of CdSe/CdS QDRs to migrate toward the droplet contact line (with respect to toluene). In addition, water presents a reduced Marangoni flow[45, 46] compared to organic solvents, which is advantageous since the Marangoni flow induces convection in the droplet that can interfere with the deposition of material at the contact line.

The emission spectrum in Figure 4c shows single-mode laser emission from the coffee-ring under fs-excitation (excitation fluence of 15 $\mu$J/cm$^2$). Laser emission is observed at 632 nm with a FWHM of 0.79 nm (see Figure SI3), close to the resolution limit of our spectrometer. A characteristic lasing power dependence is shown in Figure 4d, with a threshold of about 11 $\mu$J/cm$^2$, about 20 times lower than for previously reported CdSe/CdS coffee-rings fabricated from organic solvents.[8] The lower threshold can be tentatively explained by the higher volume fraction of CdSe/CdS per QDR granted by the small MPA surface ligands ($V_f \sim 0.4$ compared to $V_f \sim 0.25$ for ODPA). This will lead to more densely packed QDRs assemblies compared to ODPA. Additionally, the higher density will affect the total refractive index of the assembly that, combined with the slightly higher refractive index of MPA ($n_{MPA} \sim 1.49$) compared to ODPA ($n_{MPA} \sim 1.46$), can lead to an overall improvement of the reflectivity at the coffee-ring/air interface.

Single-mode lasing can occur from the coffee-rings if the free spectral range (FSR) of the cavity is larger than the bandwidth of the gain material, and the latter was evaluated as 8 nm from the FWHM of the ASE spectra of the water soluble QDRs (Figure 3f). In coffee-ring micro-lasers the resonator is formed by the cross section of the ring deposit (about 6.8 $\mu$m FWHM as obtained from the AFM profiles in Figure 4b). Assuming a Fabry-Perot cavity with length similar to the deposit FWHM, we estimate a FSR of about 14.7 nm (see SI), clearly larger than the QDRs gain bandwidth value. Even though the complex shape of the coffee ring deposit can reduce the FSR,[8] the width of 6.8 $\mu$m of the coffee-ring is small enough to sustain single-mode operation.

WILEY-VCH

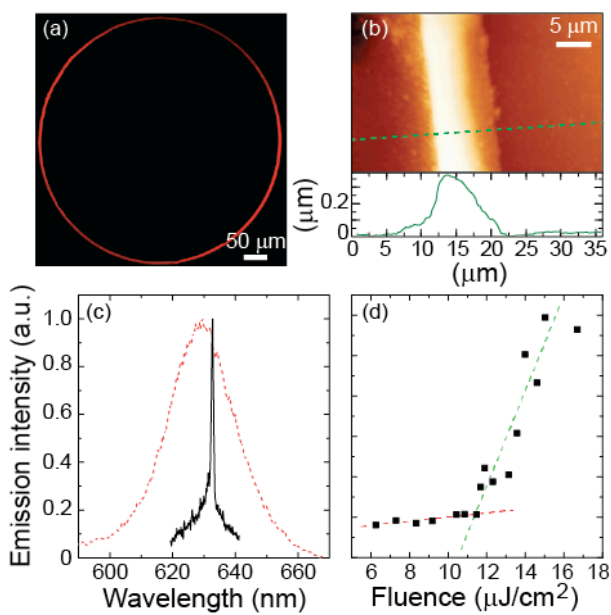

**Figure 4.** (a) Confocal microscope image of a CdSe/CdS QDRs coffee-ring. The size of these self-assembled structures can be tuned by increasing the amount of solution injected through the capillary on the substrate, or by changing the QDRs concentration in solution. Typically, coffee-ring diameters can range between 100 μm to few mm. (b) AFM image of the top-right part of the coffee-ring shown in (a). The QDRs deposit has a bottom width of about 10 μm (FWHM 6.8 $\mu$m) and a height of 370 nm. (c) Single-mode lasing (black line) from the QDR coffee-ring shown in panel (a). The lasing peak is observed at 632 nm, slightly red-shifted compared to the PL peak at 630 nm (red dashed line) with a FWHM of 0.79 nm, close to the spectral resolution of our spectrometer. (d) Emission intensity of the coffee-ring at increasing excitation fluences. A clear lasing threshold is observed at ca. 10 μJ/cm$^2$. This threshold is more than 20 times lower than previously reported.[8]

### 3. Conclusion

Using QDRs with a CdS rod with a 1.15 nm larger diameter than the CdSe core allows exchanging the surface ligands without strongly affecting the photophysics of the pristine heterostructure. By exploiting this procedure, we have demonstrated ASE and laser emission from MPA capped QDRs that are stable in water. Self-assembled coffee-ring QDR micro-lasers showed a lower lasing threshold when fabricated from water compared to organic solvents. Furthermore, the water processability of NCs greatly increases the versatility of this class of colloidal semiconductor materials. For example, by enabling the exploitation of orthogonal solvents for the fabrication of multi-layer structures (e.g. solution processable light-emitting diodes),[6, 47, 48] as well as the use of polyelectrolytes and polymer based structures for the fabrication of a variety of photonic crystals (e.g. synthetic polystyrene opals[49-52] and polymer-based flexible microcavities).[23] The versatile material design route proposed here could be extended to CdSe/CdS nanocrystals of different shapes as core-shell, dot-in-rods, and nanoplatelets, enabling their use in water solution.

## 4. Experimental Section

*Sample preparation:* CdSe/CdS QDRs were synthesized as previously reported. [15] Water-solubilization of the CdSe/CdS QDRs was carried out following the procedure in ref. [26]: 1-2 mL of 0.1 M MPA and 0.12 M KOH methanol solution were added to 100 μL of toluene solution of ODPA capped dot-in-rods with concentration of 5 - 10 μM. The mixture was left to stir for 20 minutes to complete the ligand exchange reaction and followed by centrifugation and purification through precipitation using propanol. The QDRs with MPA capping were then dispersed in deionized water.

Thin-films of organic and water soluble CdSe/CdS QDRs were obtained by drop-casting solutions with a concentration of 5 - 10 μM onto a soda-lime glass slide and dried in solvent vapor saturated atmosphere at room temperature to obtain densely packed, uniform films.

*Sample characterization.* Optical absorption spectra were collected from dilute solutions of both ODPA (dispersed in chloroform) and MPA (dispersed in water) capped QDRs using a Cary 5000 UV-Vis spectrophotometer from Agilent technologies. PL studies were carried out with an Edinburgh Instruments fluorescence spectrometer (FLS920), which included a Xenon lamp with monochromator for steady-state PL excitation, and a time-correlated single photon counting unit coupled with a pulsed laser diode (λ = 405 nm, pulse width = 50 ps) for time-resolved PL studies. A calibrated integrating sphere was used for PLQY measurements. CdSe/CdS QDRs solutions for PLQY measurements were prepared in quartz cuvettes and carefully diluted to 0.1 optical density at the excitation wavelength (λ = 450 nm).

*ASE and laser emission measurements*: CdSe/CdS QDRs films were excited with λ = 400 nm using an amplified Ti:Sapphire laser (Coherent Legend Elite seeded by a Ti:Sapphire fs laser) with a 70 fs pulse (FWHM) and a repetition rate of 1 kHz. The ASE measurements were performed by focusing the beam with a cylindrical lens onto the sample. The resulting excitation stripe dimensions were 110 x 4000 $\mu$m. Laser emission measurements on the coffee-ring were conducted using a spherical lens focusing to a spot with a 1 mm radius. All spectra were collected with a Hamamatsu Photonics streak camera, with a 7.5 cm focal length lens. ASE spectra were collected at ~90º with respect to the excitation beam.

*Self-assembled micro-laser fabrication:* Micro-lasers were fabricated by depositing 10 nL of a diluted CdSe/CdS QDRs water-solution (concentration ca. 0.1 μM) on a glass substrate. Deposition was carried out using a capillary jet technique as previously reported.[7] The apparatus consisted of an Eppendorf FemtoJet coupled with an inverted microscope to monitor the deposition process. The FemtoJet system consisted of a capillary tube (internal diameter of 0.5 μm) connected to a compressor, which controls the flow of the solution via pressure and time of the injection pulse (typical values were 1000 hPa for 0.1–1 s). The capillary tube apex was brought in close proximity to the substrate for the deposition (less than 300 μm). The substrate was previously cleaned in acetone and isopropanol in ultrasonic baths and blow dried by nitrogen. The obtained micro-lasers were then characterized by fluorescence imaging (NIKON A1 confocal microscope system) and non-contact atomic force microscopy (Nanosurf Easyscan).

**Supporting Information**

Supporting Information is available from the Wiley Online Library.

**Acknowledgements**

The research leading to these results has received funding from the CARIPLO foundation through the project "NANOCRYSLAS", and from the European Union Horizon2020 Programme under grant agreement n° 604391 FET Flagship GRAPHENE. V.L. gratefully acknowledges support from a Marie Curie Intra European Fellowship within the 7[th] European Community Framework Programme under the grant agreement n. 301100, project "LOTOCON". We thank Francesco De Donato for support in the synthesis of the CdSe/CdS dot-in-rods and the TEM images, and Marco Scotto for valuable technical assistance.

**In optoelectronic device fabrication, water solubility of nanocrystal materials is advantageous for simple and environmentally friendly processing routes.** Towards light emitting devices, highly luminescent core-shell quantum dot-in-rods (QDRs) have demonstrated excellent performance as emitters. An increase in shell thickness allows the transfer of CdSe/CdS QDRs into aqueous solution without reducing their emission, and simple drop deposition leads to single mode microlasers with low threshold.

**Keywords:** Colloidal nanocrystals, quantum dot-in-rods, water soluble, amplified spontaneous emission, laser

Dr. F. Di Stasio, Dr. J.Q. Grim, Dr. V. Lesnyak, P. Rastogi, Prof. L. Manna, Dr. I. Moreels, Prof. R. Krahne

**Single-mode lasing from colloidal water-soluble CdSe/CdS quantum dot-in-rods**

Table of contents figure:

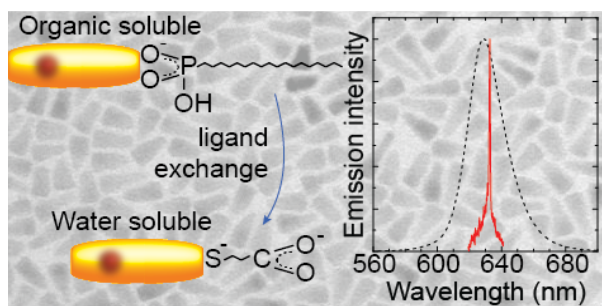

**Supporting information**

**Single-mode lasing from colloidal water-soluble CdSe/CdS quantum dot-in-rods**

*Francesco Di Stasio, Joel Q. Grim, Vladimir Lesnyak, Prachi Rastogi, Liberato Manna, Iwan*

*Moreels and Roman Krahne*

Istituto Italiano di Tecnologia, Via Morego 30, IT-16163 Genoa, Italy

**Transmission electron microscope images of ODPA and MPA capped quantum dot-in-rods:**

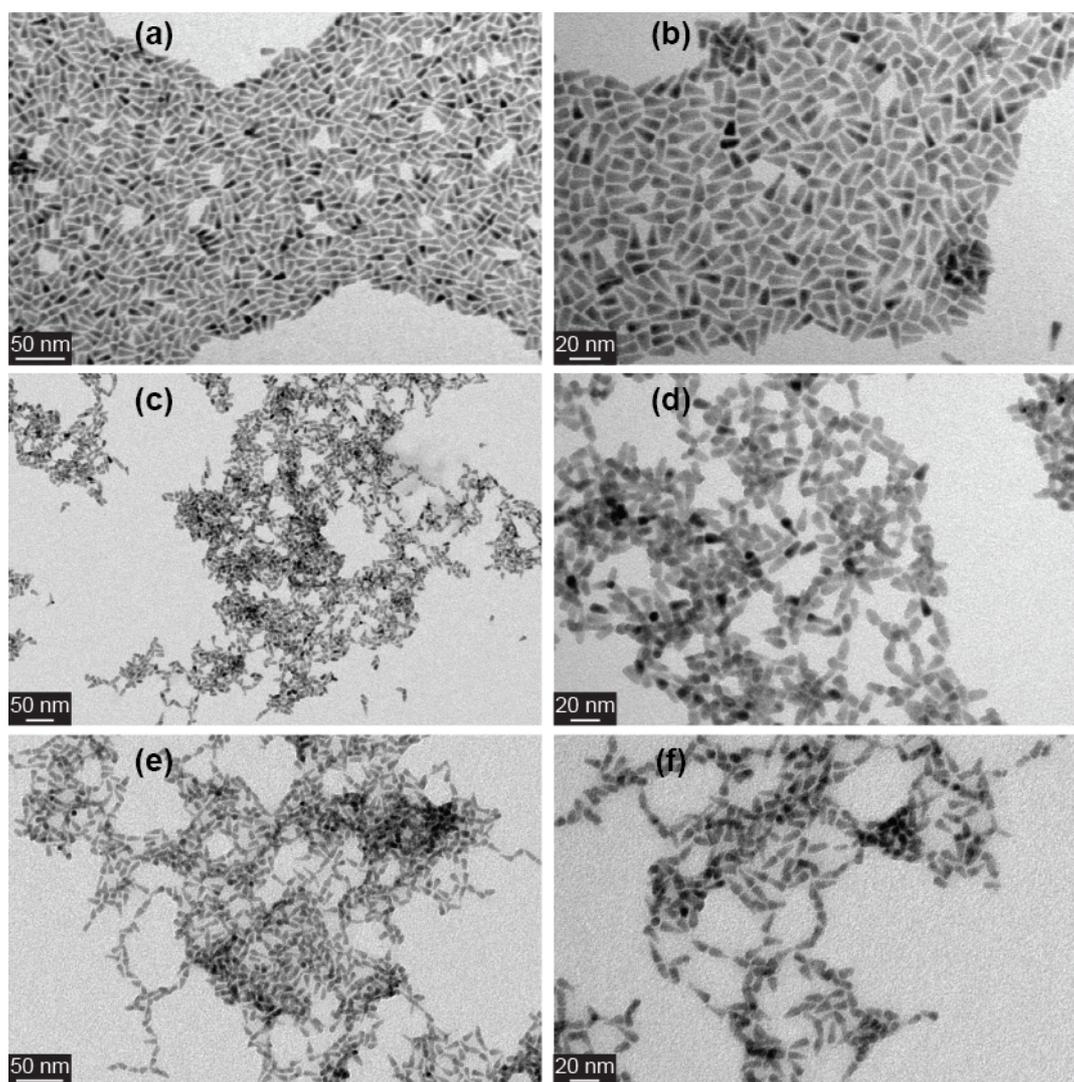

**Figure SI1** Transmission electron microscope (TEM) images at different magnification of ODPA capped (a and b) and MPA capped (c – f) quantum dot-in-rods (QDRs).

**Volume fraction per QDR estimation:**

Assuming the shape of a single QDR to be a cylinder, we calculated the volume of the CdSe/CdS using the sizes estimated from the TEM images (length $L$ =15.8 nm and diameter $d$ = 6.6 nm, see Fig. 1):

$$V_{CdSe-CdS} = \pi L \left(\frac{d}{2}\right)^2 = 540.5\ nm^3 \qquad (1)$$

Similarly, we estimated the total volume of a QDR by adding to the CdSe/CdS diameter and length the average inter-QDRs distance measured from the TEM images (see Fig. 2 and Fig. SI1), namely 2.5 nm ($d_{ODPA}$) and 1.6 nm ($d_{MPA}$) for ODPA and MPA capped QDRs respectively:

$$V_{QDR-ODPA} = \pi (L + d_{ODPA}) \left(\frac{d}{2} + d_{ODPA}\right)^2 = 1933.9\ nm^3 \qquad (2)$$

$$V_{QDR-MPA} = \pi (L + d_{MPA}) \left(\frac{d}{2} + d_{MPA}\right)^2 = 1304.7\ nm^3 \qquad (3)$$

Knowing the total volume of a single QDR ($V_{QDR}$) and the volume of the CdSe/CdS component ($V_{CdSe-CdS}$), it is possible to estimate the volume fraction ($V_f$) of the CdSe/CdS with respect to the total volume:

$$V_{f-ODPA} = V_{CdSe-CdS}/V_{QDR-ODPA} = 0.27 \qquad (4)$$

$$V_{f-MPA} = V_{CdSe-CdS}/V_{QDR-MPA} = 0.41 \qquad (5)$$

**Maxwell-Garnett effective medium theory:**

The local field factor ($f_{LF}$) is defined as:

$$f_{LF} = 3\varepsilon_s/(2\varepsilon_s + \varepsilon_{QDRs}) \qquad (6)$$

Where $\varepsilon_s$ is the dielectric constant of the solvent and $\varepsilon_{QDRs}$ is the dielectric constant of the QDRs. The local field factor $f_{LF}$ is related to the oscillator strength of the optical transition ($\varphi$) and the PL radiative decay rate ($K_R$) through Fermi's golden rule:[1, 2]

$$K_R = \frac{e^2}{2\pi\varepsilon_0 c^3 m_e}\varphi\sqrt{\varepsilon_s}|f_{LF}|^2\omega^2 \qquad (7)$$

Using the bulk CdS dielectric constant ($\varepsilon = 7.69$) at the band edge, we can estimate a $|f_{LF-ODPA}|^2 = 0.2784\ |f_{LF}|^2$ for ODPA capped QDRs in chloroform ($\varepsilon = 2.08$) and $|f_{LF-MPA}|^2 = 0.2242$ a value of 0.2242 for MPA capped QDRs in water ($\varepsilon = 1.44$). Assuming no changes in the excited state geometry[3] (affecting $\varphi$) and similar emission frequency ($\omega$), the ratio between the PL radiative rates before and after the ligand exchange can be simplified to:

$$\frac{K_{R-ODPA}}{K_{R-MPA}} = \frac{\sqrt{\varepsilon_{S-CHCl3}}|f_{LF-ODPA}|^2}{\sqrt{\varepsilon_{S-H2O}}|f_{LF-MPA}|^2} = 1.346 \qquad (8)$$

The radiative rates ratio reveals that $K_{R-MPA}$ is $\sim 35\%$ lower than $K_{R-ODPA}$ due to differences in $f_{LF}$. This value is in good agreement with the $\sim 42\%$ PL lifetime increase for MPA ($\tau_{MPA} = 20\ ns$) with respect to ODPA ($\tau_{ODPA} = 14\ ns$) capped QDRs. These considerations suggest that the measured PL lifetime variation is predominantly due to the variation of $\varepsilon_s$ and consequently $|f_{LF}|^2$.

**Optical properties of water-soluble MPA capped CdSe/CdS quantum dot-in-rods (QDRs) with similar CdSe and CdS diameters:**

Applying the MPA ligand exchange method to CdSe/CdS QDRs with CdS rods with similar diameter to the CdSe core (see Fig. SI2) leads to the formation of additional PL quenching channels that strongly affect the PLQY of the system (see Fig. SI2c). PLQY decreases from $50 \pm 5\%$ to $10 \pm 1\%$ after the ligand exchange reaction with concomitant decrease in PL lifetime from 20 ns for ODPA capped QDRs to 8 ns for MPA capped ones, as shown in Fig. SI1d, which is related to the marked increase in the non-radiative rate.

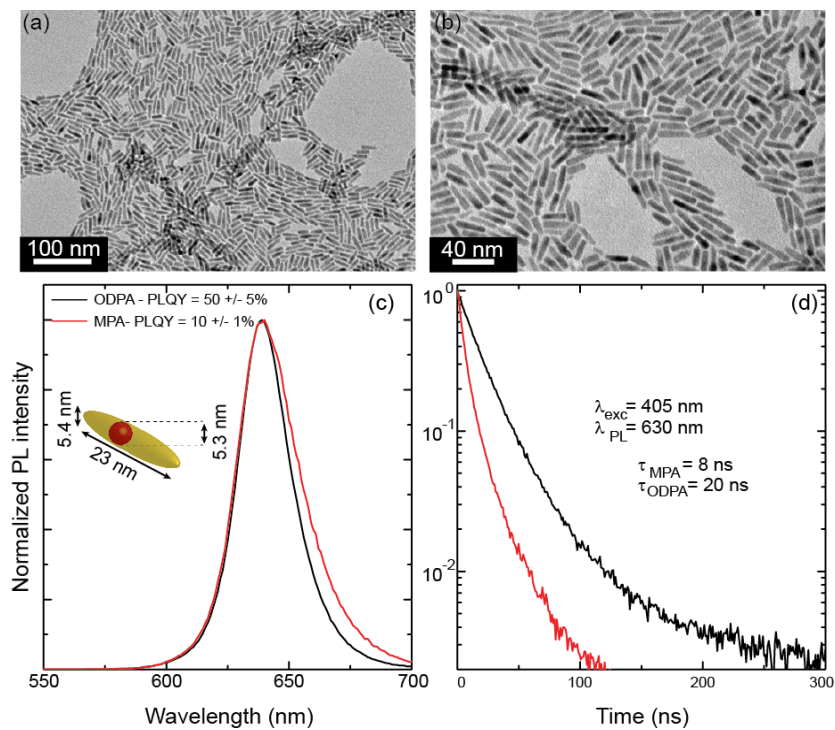

**Figure SI2** TEM images (a, b) at different magnifications of ODPA capped QDRs.(c) PL spectra of the QDRs before (ODPA capped) and after (MPA capped) the ligand-exchange reaction. A small PL broadening is observed. (Inset: scheme of a QDR used with average dimensions determined via TEM analysis). (d) PL time decay measured at the PL peak for ODPA and MPA capped QDRs.

**Fabry-Perot Cavity free spectral range for the self-assembled micro-laser:**

Modelling the coffee ring structure as a standard Fabry-Perot cavity, the free spectral range (FSR) can be defined as:

$$FSR = \frac{\lambda^2}{2nL} \qquad (4)$$

where $L$ is the cavity width, $n$ the refractive index, and $\lambda$ the emission wavelength. Using a cavity width equal to 6.8 μm ( as the full-width-half-maximum of the deposit), we can calculate an FSR of 14.7 nm (assuming a refractive index $n$ = 2 for the MPA capped QDRs).

**Single-Mode lasing from a self-assembled micro-laser:**

WILEY-VCH

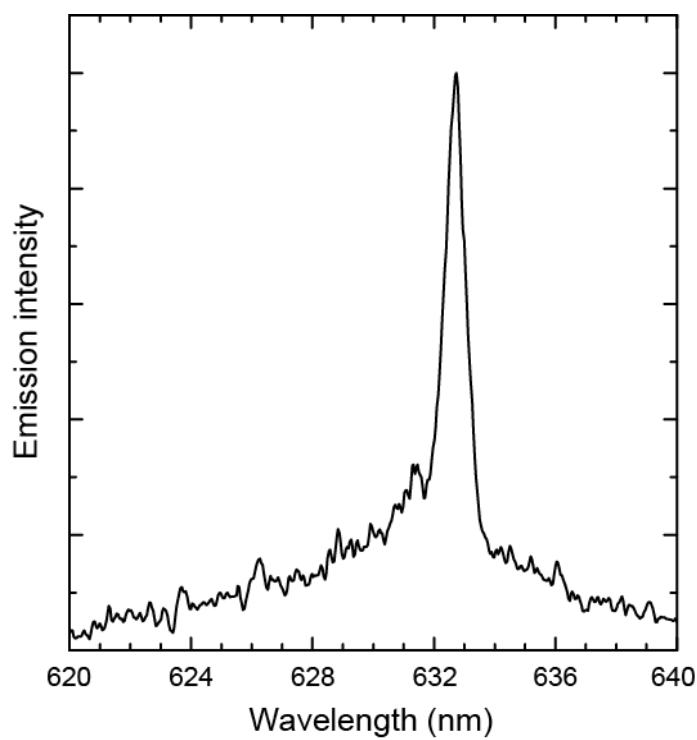

**Figure SI3** Laser emission from an MPA capped QDRs coffee-ring under fs-pulsed excitation at a pump fluence of 15 μJ/cm$^2$. The full-width-half-maximum of the lasing peak is 0.79 nm, close to the spectral resolution of our spectrometer.

**Polarized optical microscope images of a self-assembled micro-laser:**

Previously reported self-assembled micro-laser fabricated with ODPA-capped QDRs consisted of coffee-rings in which the nanocrystals were mainly aligned tangential to the ring perimeter. This alignment could be detected by polarized microscopy, since it results in an anisotropy of the refractive index of the QDRs deposit.[4-6] Surprisingly, we did not observe such alignment for MPA-capped QDRs, where lack of birefringence indicates a disordered structure.

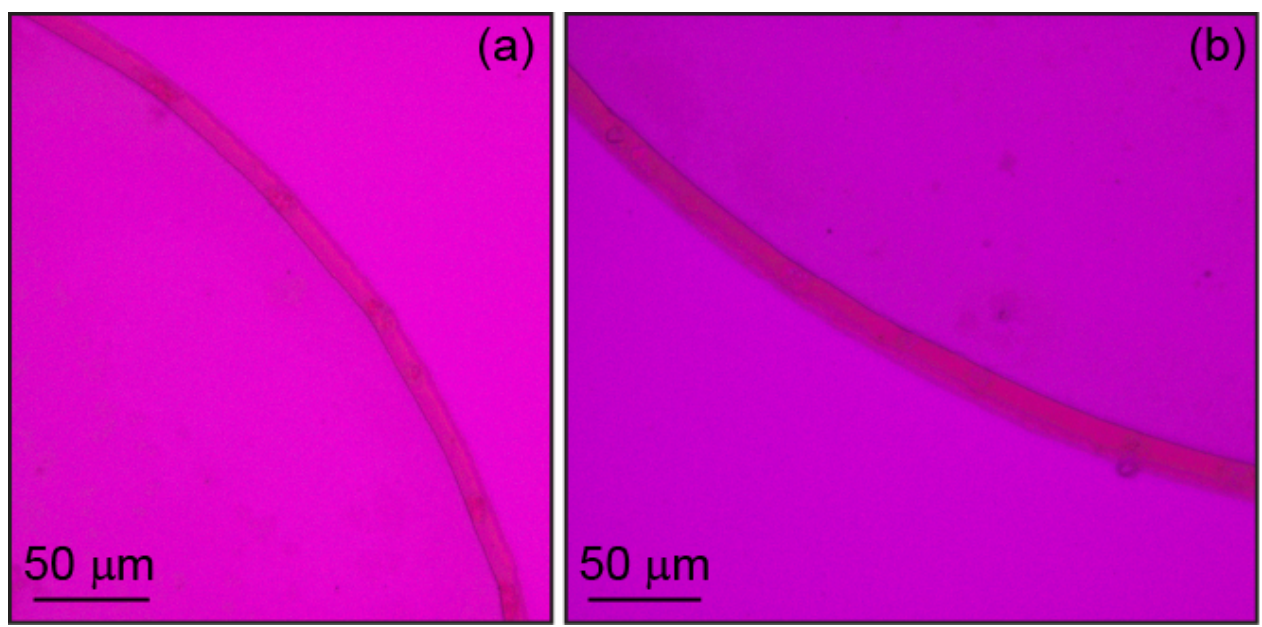

**Figure SI4** Polarized optical microscope images of a typical self-assembled micro-laser fabricated with MPA-capped QDRs. The sample in (b) was rotated by 90° with respect to (a). The images were recorded with a retardation plate at 530 nm oriented at 45° with respect to the axes of the crossed polarizers. Therefore, the magenta color corresponds to an isotropic refractive index, indicating no specific alignment order in the QDRs assembly.